\documentclass[12pt,a4paper]{article}

\usepackage[utf8]{inputenc}
\usepackage[T1]{fontenc}
\usepackage{amsmath,amssymb}
\usepackage{graphicx}
\usepackage[margin=2.5cm]{geometry}
\usepackage{booktabs}
\usepackage{array}
\usepackage{longtable}
\usepackage{multirow}
\usepackage{xcolor}
\usepackage{hyperref}
\usepackage{natbib}
\usepackage{caption}
\usepackage{subcaption}
\usepackage{tabularx}
\usepackage{makecell}

\hypersetup{colorlinks=true,linkcolor=blue,citecolor=blue,urlcolor=blue}

\title{A Practical Guide to Interpret a Randomized Controlled Trial:\\ Underpowered $\neq$ Inconclusive $\neq$ Negative $\neq$ Neutral}

\author{
Ibrahim Halil Tanboga, MD, PhD\\
\textit{Department of Biostatistics and Cardiology, Nisantasi University Medical School, Istanbul, Turkey}\\
\texttt{haliltanboga@gmail.com}
}

\date{}

\begin{document}
\maketitle

\begin{abstract}
The most dangerous error in clinical trial interpretation is equating $p > 0.05$ with ``no effect.'' This review provides a practical, algorithm-based framework for classifying randomized controlled trial (RCT) results into six distinct categories---positive, imprecise~(+), neutral, inconclusive, negative, and harmful---using confidence interval (CI) position relative to the minimal clinically important difference (MCID) as the primary tool, augmented by Bayesian posterior probabilities. We demonstrate that the same $p > 0.05$ result can represent three fundamentally different conclusions (inconclusive, negative, or neutral), show how Bayesian reanalysis can rescue benefit signals missed by frequentist thresholds, and illustrate the framework with real-world examples from critical care and cardiology trials. The framework synthesizes guidance from Altman, Harrell, Pocock, Zampieri, the ASA, and ICH~E9 into a single coherent decision algorithm.
\end{abstract}

\tableofcontents
\newpage

%% ============================================================
\section{Introduction: Why $p$-values Alone Are Insufficient}
%% ============================================================

The single most dangerous error in clinical trial interpretation is equating $p > 0.05$ with ``no effect.'' Altman and Bland formalized this in \textit{BMJ} (1995) \citep{altman1995}: ``Absence of evidence is not evidence of absence.''

The American Statistical Association (ASA, 2016) stated unequivocally that scientific conclusions should not be based solely on whether a $p$-value crosses a specific threshold \citep{wasserstein2016}. A small $p$-value does not necessarily indicate a large or important effect, and a large $p$-value does not mean the effect is absent.

Harrell argues the terms ``positive'' and ``negative'' should be abandoned entirely, replaced by continuous probability statements \citep{harrell2017}. His blog epigraph captures this: ``To avoid `false positives' do away with `positive.'\,'' On the Datamethods forum (2018), he proposed increasingly honest formulations for non-significant results:

\begin{quote}
``The most brutally honest summary of an underpowered study: `The money was spent.' Less brutal: `The presumption of no difference is not yet overcome by data.'\,''
\end{quote}

Critically, he acknowledged that even this language was ``still too `dichotomous' in that it was written assuming we would have different language for `positive' vs.\ `negative' studies. This implies a threshold for `positive' which is what we're trying to get away from.''

On underpowered trials specifically, Harrell is withering: ``Underpowered trials are worse than no trials. Because used as evidence they can mislead especially those who aren't cognizant of power \& errors.''

The only reliable way to classify an RCT result is to examine the position of the 95\% confidence interval (CI) relative to the pre-specified MCID ($\delta$) and the null value. The $p$-value alone cannot distinguish positive, negative, neutral, or inconclusive outcomes. When available, Bayesian posterior probabilities resolve the remaining ambiguity that even CI~+~MCID cannot fully address.

%% ============================================================
\section{The Complete Decision Algorithm}
%% ============================================================

This algorithm has two parallel tracks: a \textbf{frequentist track} (CI~+~MCID) and a \textbf{Bayesian track} (posterior probabilities). The frequentist track is sufficient for most cases; the Bayesian track adds decisive value when the $p$-value falls near 0.05 or when the neutral vs.\ negative distinction matters clinically.

% ---- Track A ----
\subsection{Track A --- Frequentist (CI + MCID)}

\paragraph{Step 1 --- Define effect measure, null value, and MCID ($\delta$).}
The effect measure may be HR, RR, OR, mean difference, or ARD. Null value: HR/RR/OR~$= 1.0$, mean difference/ARD~$= 0$.

\textbf{What is the MCID?} The Minimal Clinically Important Difference (MCID) is the smallest treatment effect that would be considered clinically meaningful to patients and clinicians---the threshold below which a statistically significant result would not change clinical practice. The MCID is \emph{not} a post-hoc concept: it must be pre-specified in the trial's statistical analysis plan, typically in the ``Statistical Methods'' section of the protocol, \emph{before} data collection begins. It is determined by a combination of clinical judgment, prior evidence, and patient-centered outcome data \citep{mcglothlin2013}.

For example, in a cardiovascular mortality trial, a hazard ratio (HR) of 0.95 (5\% relative reduction) would generally not change practice, whereas HR~$\leq 0.80$ ($\geq$20\% relative reduction) would be considered a clinically meaningful benefit. Thus, the MCID-benefit threshold would be set at $\delta = 0.80$. Similarly, a harm threshold might be set at HR~$\geq 1.25$ (25\% relative increase in the endpoint). In heart failure trials, where baseline event rates are high, even a 15\% relative reduction (HR~$\leq 0.85$) may qualify as the MCID. In oncology, where therapies carry significant toxicity, MCID thresholds tend to be more demanding. The key point is that the MCID anchors the entire classification framework: without it, a CI can tell you about precision but cannot tell you about clinical relevance.

\paragraph{Step 2 --- Does the 95\% CI exclude the null?}
If \textbf{YES} ($p < 0.05$):
\begin{itemize}
  \item[2a.] Entire CI beyond MCID-benefit $\rightarrow$ \textsc{Positive}
  \item[2b.] CI crosses MCID (includes effects both above and below $\delta$) $\rightarrow$ \textsc{Imprecise~(+)} --- significant but magnitude uncertain
  \item[2c.] Entire CI in harm zone beyond MCID-harm $\rightarrow$ \textsc{Harmful}
\end{itemize}
If \textbf{NO} ($p \geq 0.05$) $\rightarrow$ go to Step~3.

\paragraph{Step 3 --- How wide is the CI relative to MCID zones?}
\begin{itemize}
  \item Narrow CI within $[-\delta, +\delta]$ indifference zone --- both MCID-benefit and MCID-harm excluded $\rightarrow$ \textsc{Neutral} (precise null)
  \item Narrow CI excludes MCID-benefit but \emph{not} MCID-harm --- clinically meaningful benefit ruled out $\rightarrow$ \textsc{Negative} (benefit excluded)
  \item Wide CI crosses MCID-benefit and/or MCID-harm --- includes both clinically meaningful benefit and harm $\rightarrow$ \textsc{Inconclusive}
\end{itemize}

\paragraph{Step 4 --- NEVER compute post-hoc power.}
CI already contains all precision information. Post-hoc power $= f(p\text{-value}) =$ zero additional information.

% ---- Figure 1: Flowchart ----
\begin{figure}[!htbp]
\centering
\includegraphics[width=0.95\textwidth]{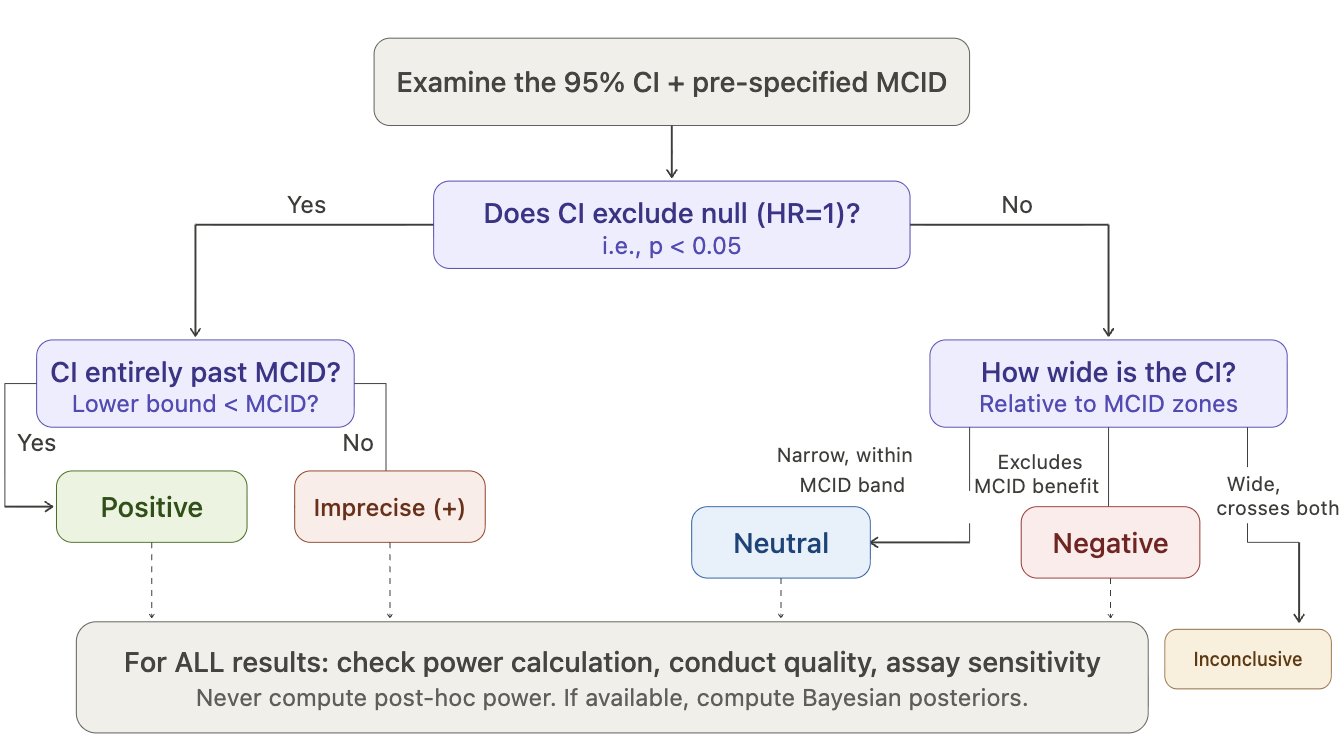}
\caption{\textbf{CI~+~MCID Decision Flowchart.} The frequentist classification algorithm. First determine whether the 95\% CI excludes the null; then classify by CI width relative to MCID zones. Post-hoc power should never be computed.}
\label{fig:flowchart}
\end{figure}

% ---- Track B ----
\subsection{Track B --- Bayesian (Zampieri/Harrell Framework)}

\paragraph{Step 5 --- Define priors (minimum 3 per Zampieri et al.\ AJRCCM 2021).}
Select belief strength (weak/moderate/strong) based on prior evidence per Zampieri Table~1 \citep{zampieri2021}. See Table~\ref{tab:priors}.

% ==== TABLE 1: Prior Types ====
\begin{table}[!htbp]
\centering
\caption{\textbf{Prior specifications for Bayesian reanalysis} (adapted from Zampieri et al.\ 2021).}
\label{tab:priors}
\begin{tabular}{@{}llll@{}}
\toprule
\textbf{Prior type} & \textbf{Center} & \textbf{Distribution} & \textbf{Constraint} \\
\midrule
Skeptical (null-centered) & OR $= 1$ & $\mathcal{N}(0,\; 0.355)$ & Moderate strength \\
Optimistic (benefit-centered) & Expected benefit & $\mathcal{N}(-\delta,\; \sigma)$ & $\Pr(\text{harm}) \geq 15\%$ \\
Pessimistic (harm-centered) & Expected harm & $\mathcal{N}(+\delta,\; \sigma)$ & $\Pr(\text{benefit}) \geq 15\%$ \\
Data-derived (meta-analysis) & Pooled estimate & $\mathcal{N}(\hat\mu,\; \hat\sigma)$ & Optional \\
\bottomrule
\end{tabular}
\end{table}

\paragraph{Step 6 --- Compute 3 posterior metrics for each prior.}
See Table~\ref{tab:metrics} for metric definitions.

\paragraph{Step 7 --- Classify by dominant metric.}
See Table~\ref{tab:bayesian_criteria} for classification rules.

\paragraph{Step 8 --- Sensitivity check.}
Compute $I^2$ across prior results. If $I^2 < 0.20$, conclusion is robust to prior choice --- the data dominate.

\textbf{Key principle:} if the verdict is the same across skeptical, optimistic, and pessimistic priors $\rightarrow$ the data are talking, not the prior.

%% ============================================================
\section{The CI + MCID Classification}
%% ============================================================

\begin{figure}[!htbp]
\centering
\includegraphics[width=0.95\textwidth]{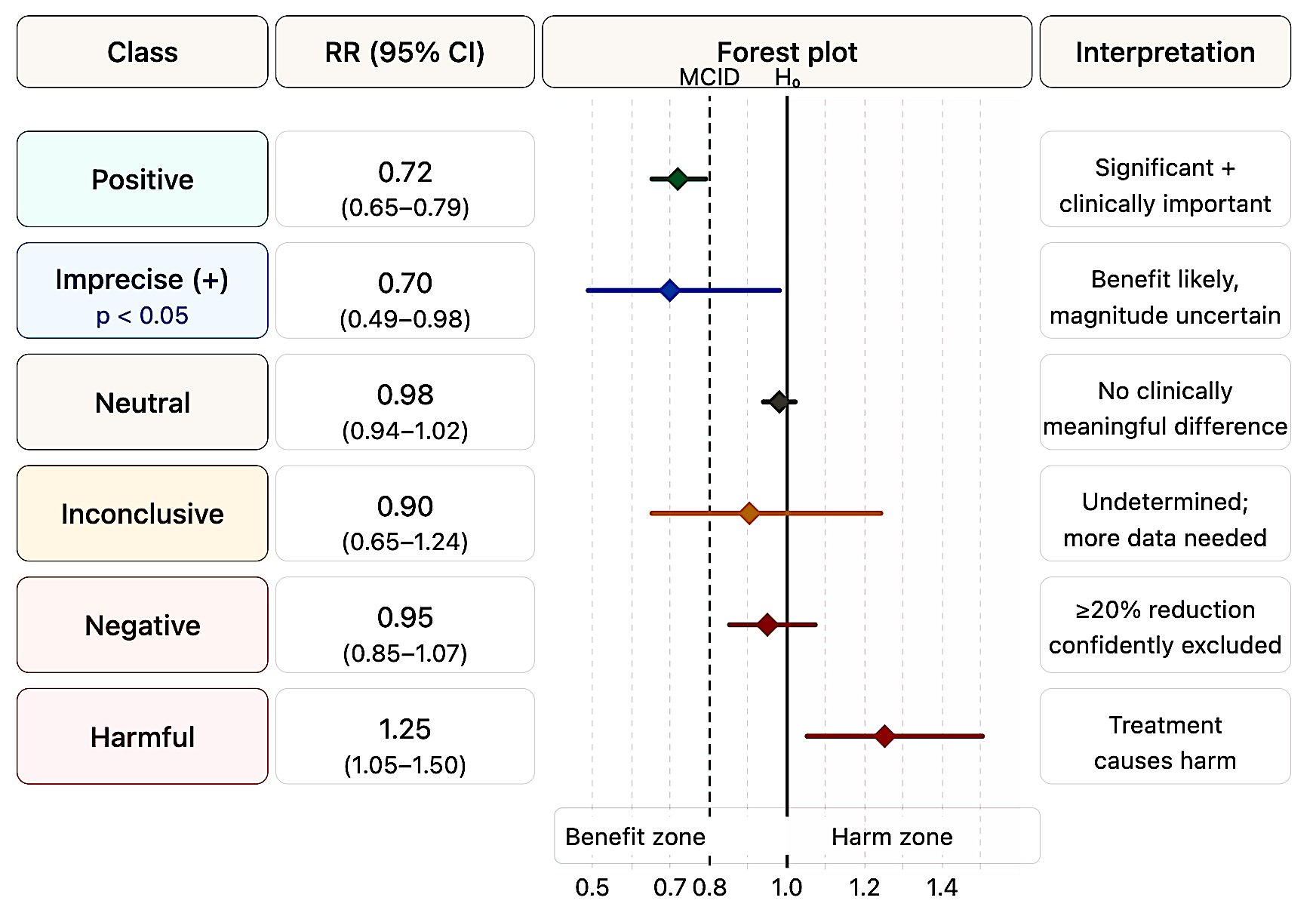}
\caption{\textbf{Six-class forest plot.} Each row illustrates a prototypical CI position. The dashed lines mark the MCID for benefit (left) and harm (right); the solid line marks the null ($\text{HR}=1.0$). Classification depends on CI position relative to these thresholds, not on the $p$-value alone.}
\label{fig:forest}
\end{figure}

Figure~\ref{fig:forest} presents a forest plot illustrating the six possible trial classifications based on CI position relative to the null and MCID thresholds.

% ---- Figure 2: Forest plot ----

%% ============================================================
\subsection{How Leading Authorities' Terminologies Compare and Conflict}
\label{sec:terminology}
%% ============================================================

The terms ``positive,'' ``negative,'' ``neutral,'' and ``inconclusive'' are used inconsistently across the literature. Table~\ref{tab:authorities} maps each authority's preferred terms onto the six categories in this framework.

% ==== TABLE 4: Authority Terminology (longtable for multi-page) ====
{\small
\renewcommand{\arraystretch}{1.2}
\begin{longtable}{@{}p{2.2cm}p{2cm}p{2.5cm}p{2.5cm}p{2.3cm}p{2cm}@{}}
\caption{\textbf{Terminology comparison across leading authorities.} Each cell shows how a given authority defines or uses the corresponding term. C = Classical/frequentist; B = Bayesian; F = Framework/regulatory.}
\label{tab:authorities} \\
\toprule
\textbf{Authority} & \textbf{Positive} & \textbf{Negative} & \textbf{Inconclusive} & \textbf{Under\-powered} & \textbf{Neutral} \\
\midrule
\endfirsthead
\toprule
\textbf{Authority} & \textbf{Positive} & \textbf{Negative} & \textbf{Inconclusive} & \textbf{Under\-powered} & \textbf{Neutral} \\
\midrule
\endhead
\midrule
\multicolumn{6}{r}{\textit{Continued on next page}} \\
\endfoot
\bottomrule
\endlastfoot
Altman \& Bland (C) & Not addressed & Narrow CI excluding meaningful effects; label should be abandoned & Default for most non-significant results & Design deficiency leading to inconclusiveness & Not distinguished from negative \\
\midrule
Freiman et al.\ (C) & Not addressed & Requires $\geq$90\% power + CI excluding meaningful effects & Implicit --- trials haven't received `a fair test' & 94\% of `negative' trials were underpowered & Not used \\
\midrule
Harrell (B) & Rejects term --- use Pr(benefit) & `Absence of evidence $\neq$ evidence of absence' & Prefers `uninformative' or `the money was spent' & `Worse than no trials' & Compute Pr(similarity) via Bayesian model \\
\midrule
Pocock \& Stone (F) & Addressed in companion paper & 12-question evaluation required & Explicitly used (e.g., PRoactive) & Question \#2 in 12-question framework & Not a primary term \\
\midrule
Hawkins \& Samuels (C) & Not addressed & Interval A: CI within MCID boundaries & Interval B or C (spanning both) & Post-hoc power `irrelevant once completed' & Subsumed under `negative or neutral' \\
\midrule
Zampieri et al.\ (B) & High Pr(benefit) across priors & High Pr(harm) across priors & `Indeterminate' --- sensitive to prior choice & Bypassed by Bayesian framework & High ROPE probability \\
\midrule
Hoenig \& Heisey (C) & Not addressed & Requires equivalence testing (TOST) & Wide CI including null and meaningful effects & Post-hoc power is a fallacy & Demonstrated via equivalence testing \\
\midrule
Gelman \& Carlin (B) & May be Type S/M corrupted & Not addressed directly & Low-power = low information & Corrupts both `positive' AND `negative' & Not addressed \\
\midrule
ASA (2016, 2019) (F) & `Don't say statistically significant' & `Large $p$ does not imply evidence for null' & Implicit --- `a conclusion does not address' & Not specifically addressed & Not addressed \\
\midrule
Goodman (B) & $p\!<\!0.05$ gives $\leq$13\% posterior null prob.\ & `Nonsignificant $\neq$ no effect' & Default state --- external evidence required & Not specifically used & Not used \\
\midrule
Senn (C) & Defends $p$-values as evidence gradients & `Not found good evidence for a real effect' & Default for non-significant results & Depends on pre-specified $\delta$ & Not used as distinct term \\
\midrule
ICH E9 (F) & Not classified by terminology & Cannot conclude from non-significant superiority & Implicit --- non-significant $\neq$ equivalent & Pre-specified power requirements & Requires formal equivalence testing \\
\midrule
Campbell \& Gustafson CET (F) & NHST rejects null & Equivalence test shows effect within $\delta$ & Neither test rejects --- explicit third category & Design property leading to inconclusive & Subsumed under `negative' if within $\delta$ \\
\end{longtable}
}

Three points of notable divergence:

\paragraph{``Inconclusive'' vs ``indeterminate'':} Harrell and Zampieri's group prefer ``indeterminate'' or ``uninformative'' --- reflecting their view that even ``inconclusive'' implies more structure than the data warrant.

\paragraph{``Neutral'' is the most inconsistently used term:} Hawkins and Samuels bundle it with ``negative,'' Zampieri et al.\ map it to high ROPE probability, and most others avoid it entirely. The unique contribution of Bayesian analysis is that only $\Pr(\text{ROPE})$ can formally operationalize ``neutral.''

\paragraph{``Underpowered'' corrupts all categories:} Altman, Bland, Hawkins, and Samuels treat underpowering as a \emph{design} property that causes inconclusiveness (a \emph{result} property). Gelman and Carlin go further: underpowering corrupts \emph{all} result categories --- including ostensibly ``positive'' findings via Type~M/S errors.

Despite terminological variation, every authority converges on a single core principle: \textbf{a non-significant $p$-value cannot, by itself, classify a trial as ``negative.''}

%% ============================================================
\section{Term Definitions and Diagnostic Criteria}
%% ============================================================

\subsection{Underpowered}

\textbf{Definition:} The study's sample size or event count was insufficient to reliably detect the pre-specified MCID at $>$80\% power. Applies to both $p < 0.05$ and $p > 0.05$ results.

\textbf{Diagnostic criteria:}
\begin{itemize}
  \item CI includes both the null value and the MCID (wide CI)
  \item Protocol power calculation target (effect size or event count) was not achieved
  \item If $p < 0.05$: high risk of Type~M (magnitude exaggeration) and Type~S (sign error) per Gelman \& Carlin (2014) \citep{gelman2014}
\end{itemize}

\textbf{The core insight --- same drug, same truth, different sample sizes:} Two trials can find the same point estimate (e.g., HR~0.82) but tell completely different stories:
\begin{itemize}
  \item $n = 500$ (underpowered): CI 0.55--1.18 $\rightarrow$ spans 48\% benefit to 18\% harm. Compatible with everything. Harrell: ``Almost nothing was learned.'' $\rightarrow$ \textsc{Inconclusive}
  \item $n = 4{,}500$ (adequate): CI 0.74--0.91 $\rightarrow$ entirely in benefit zone, past MCID. $\rightarrow$ \textsc{Positive}
\end{itemize}

CI width is proportional to $1/\sqrt{n}$. To halve the CI width, you need approximately $4\times$ the sample size (or events).

The most practical way to detect underpowering: look at the CI. If it includes both the MCID and the null --- regardless of the $p$-value --- the trial is underpowered and the result is inconclusive.

\paragraph{The post-hoc power fallacy.} Post-hoc power $= f(p\text{-value})$; zero additional information. When $p = \alpha$, observed power is exactly 50\% regardless of sample size. The ``Power Approach Paradox'' \citep{hoenig2001}: experiments closer to significance have \emph{higher} observed power, paradoxically implying they better support the null. CONSORT 2010 and Christensen et al.\ (2024): ``Statistical power is exclusively a pre-trial concept.''

\subsection{Winner's Curse: Why Underpowered ``Positive'' Results Exaggerate}

\textbf{The mechanism:} In a low-powered study, the true effect is small, but estimates are very noisy. To pass the $p < 0.05$ threshold, the observed effect must randomly deviate far from truth. Most estimates cluster near the true (small) effect and fail to reach significance. Only the rare, grossly overestimated values cross the threshold. These are the ones that get published --- and they are ``cursed'' with an inflated effect.

\textbf{Gelman \& Carlin (2014) simulation} --- study with 6\% power \citep{gelman2014}:
\begin{itemize}
  \item Type~M error: $9.7\times$ --- the published effect is nearly 10 times the true effect
  \item Type~S error: 24\% --- 1 in 4 significant results has the wrong sign entirely
\end{itemize}

\textbf{Practical chain reaction:} Researcher takes the inflated effect (e.g., HR~0.65) and uses it for the confirmatory trial's power calculation $\rightarrow$ ``500 patients will suffice'' $\rightarrow$ but the true effect is HR~0.92, so 500 patients are grossly insufficient $\rightarrow$ confirmatory trial is ``negative'' $\rightarrow$ ``first result failed to replicate'' $\rightarrow$ the problem was never the drug, it was the inflated initial estimate.

\textbf{Empirical evidence:}
\begin{itemize}
  \item Button et al.\ (\textit{Nat Rev Neurosci}, 2013) \citep{button2013}: Underpowered studies (median power 8--31\%) inflate initial effect estimates by 25--50\%
  \item Ioannidis (\textit{PLoS Med}, 2005) \citep{ioannidis2005}: Low power + modest prior probability $\rightarrow$ a statistically significant result is more likely false than true (PPV $< 50\%$)
  \item Sidebotham \& Barlow (\textit{Anaesthesia}, 2024) \citep{sidebotham2024}: Directly warned against using small-trial effects for confirmatory power calculations
\end{itemize}

\subsection{Inconclusive}

\textbf{Definition:} The CI is too wide to permit a definitive classification --- it spans both the null and the MCID, or spans from clinically meaningful benefit to clinically meaningful harm.

\textbf{Diagnostic criteria:} $p > 0.05$ AND the 95\% CI crosses both the null and the MCID thresholds. Example: HR $= 0.90$ (CI 0.65--1.24) $\rightarrow$ both 35\% benefit and 24\% harm are possible.

Freiman et al.\ (\textit{NEJM}, 1978) \citep{freiman1978}: Of 71 ``negative'' RCTs, only 6\% had $\geq$90\% power to detect a 25\% improvement.

\textbf{Bayesian signature:} No probability dominates --- $\Pr(\text{benefit}) \sim 38\%$, $\Pr(\text{equivalence}) \sim 35\%$, $\Pr(\text{harm}) \sim 18\%$. The posterior is spread across all possibilities almost uniformly. This is the mathematical definition of ``the data haven't told us anything decisive.''

\paragraph{Conditional Equivalence Testing (CET).} Campbell \& Gustafson (\textit{PLoS ONE} 2018) \citep{campbell2018}: CET is the only framework that explicitly and formally defines all three outcome categories as part of a single testing procedure:
\begin{itemize}
  \item \textbf{Positive:} Standard NHST rejects the null (significant difference found)
  \item \textbf{Negative:} Equivalence test (TOST) demonstrates the effect lies within a pre-specified equivalence margin $\delta$
  \item \textbf{Inconclusive:} Neither test rejects --- insufficient evidence to claim either a difference or equivalence
\end{itemize}

The two-stage procedure works sequentially: standard NHST is performed first; if it fails to reject ($P > \alpha$), an equivalence test is conducted. CET requires pre-specification of an equivalence margin ($\delta$), analogous to the MCID. Simulation studies showed CET and Bayesian JZS Bayes Factor testing reach similar conclusions in many scenarios.

\textbf{Important:} 54--56\% of non-significant RCTs use ``no difference'' or ``no benefit'' in the abstract (Gates et al., 2019). The correct approach: emphasize uncertainty + report CI.

\subsection{Negative vs Neutral: The Critical Distinction}

These two terms are the most commonly confused. The distinction is precise:

\textbf{Negative} says: ``This treatment does NOT provide meaningful benefit.'' But it leaves an open question --- the CI may extend toward harm. Only the benefit side is excluded by MCID.\\
\textit{Example:} HR 0.95 (CI 0.85--1.07), MCID $= 0.80$ $\rightarrow$ Entire CI above 0.80, $\geq$20\% benefit excluded. But upper bound 1.07 means 7\% harm still possible.\\
\textit{Bayesian:} $\Pr(\text{MCID benefit}) \approx 3\%$, $\Pr(\text{ROPE}) \approx 88\%$, $\Pr(\text{harm}) \approx 20\%$.

\textbf{Neutral} says: ``The two treatments are essentially THE SAME.'' The CI is so narrow that both benefit and harm are excluded.\\
\textit{Example:} HR 0.98 (CI 0.94--1.02) $\rightarrow$ Neither 20\% benefit nor 25\% harm is possible.\\
\textit{Bayesian:} $\Pr(\text{MCID benefit}) < 1\%$, $\Pr(\text{ROPE}) \approx 97\%$, $\Pr(\text{harm}) \approx 5\%$.

\textbf{Key formula:}
\begin{align*}
\text{Negative} &= \text{benefit excluded (one-sided)} \rightarrow \text{``doesn't work''} \\
\text{Neutral}  &= \text{benefit AND harm excluded (two-sided)} \rightarrow \text{``they're the same''}
\end{align*}

Neutral is a much stronger statement. It requires a narrow CI and ideally formal confirmation through an equivalence test or Bayesian $\Pr(\text{ROPE})$. The unique power of Bayesian analysis is that only it can formally quantify $\Pr(\text{equivalence})$ --- the single number that defines neutral.

\subsection{Positive}
\textbf{Definition:} CI entirely on benefit side and beyond MCID --- both statistical and clinical significance. Verify adequate power; if low, suspect Type~M error (Gelman \& Carlin).

\subsection{Harmful}
\textbf{Definition:} CI entirely in harm zone beyond MCID-harm. Bayesian confirmation: $\Pr(\text{severe harm}) > 40\%$ across all priors including optimistic. The ART trial (JAMA 2017) is the paradigmatic example --- see Section~\ref{sec:art}.

%% ============================================================
\section{The Three Faces of $p > 0.05$}
\label{sec:threefaces}
%% ============================================================

This is the most important conceptual illustration in the entire framework. Three trials, all $p > 0.05$, all ``non-significant'' --- yet they mean completely different things:

\paragraph{Scenario A: Underpowered ($n=400$, $p = 0.18$).}
HR 0.78 (CI 0.52--1.18). CI spans 48\% benefit to 18\% harm. Compatible with everything.\\
$\rightarrow$ \textsc{Inconclusive} --- ``Almost nothing was learned'' (Harrell).

\paragraph{Scenario B: Negative ($n=8{,}000$, $p = 0.22$).}
HR 0.95 (CI 0.87--1.04). Narrow CI near null. Entire CI above MCID. Best case: only 13\% benefit.\\
$\rightarrow$ \textsc{Negative} --- ``Clinically meaningful benefit is ruled out.''

\paragraph{Scenario C: Neutral ($n=12{,}000$, $p = 0.48$).}
HR 0.98 (CI 0.93--1.03). Very narrow CI hugs null. Both benefit and harm excluded.\\
$\rightarrow$ \textsc{Neutral} --- ``No meaningful difference in either direction.''

\textbf{The tragedy:} Most papers report all three as ``no significant difference'' --- a single phrase hiding three fundamentally different conclusions. This is the error Altman warned about in 1995 and that persists in over half of published RCTs today.

% ---- Figure 3: Three Faces ----
\begin{figure}[!htbp]
\centering
\begin{subfigure}[b]{0.32\textwidth}
\includegraphics[width=\textwidth]{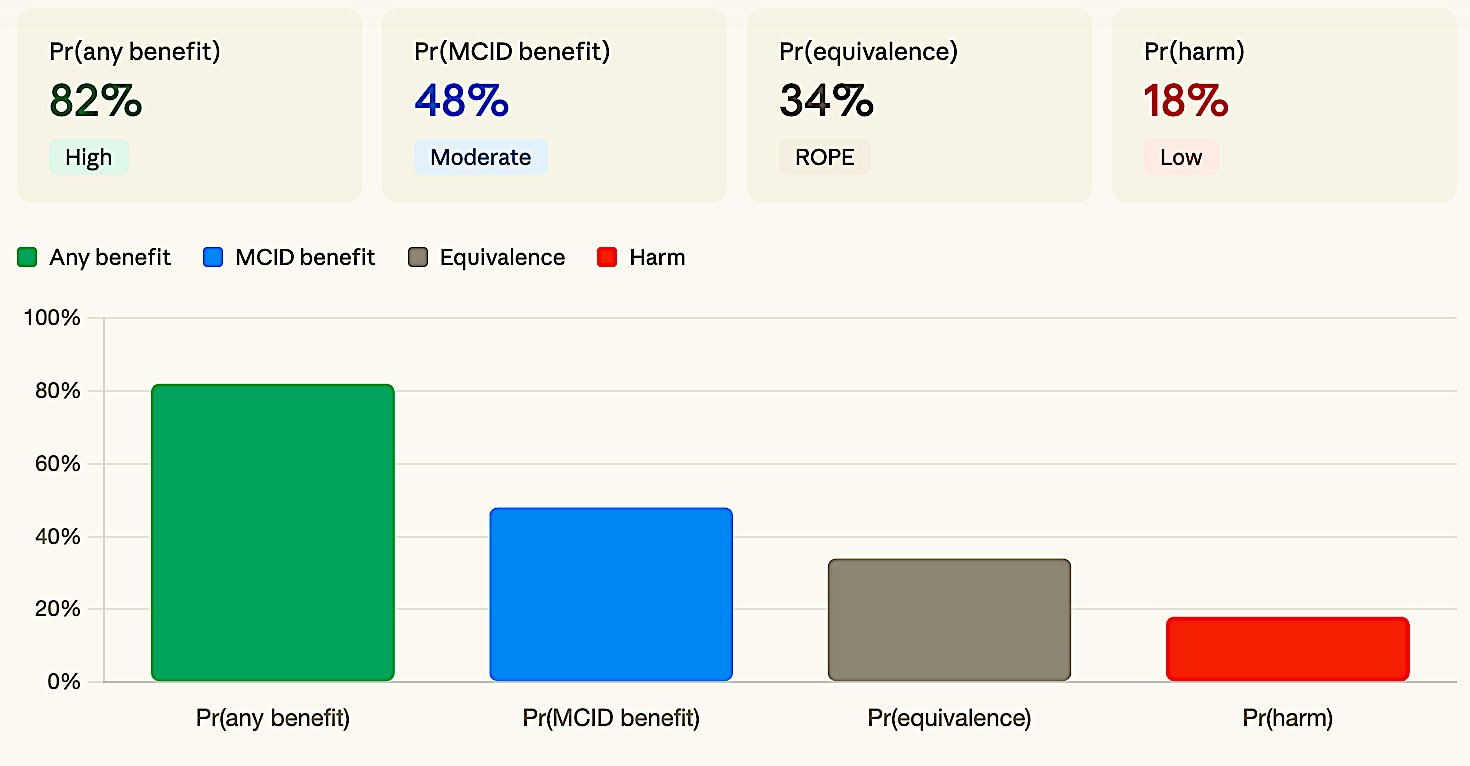}
\caption{Inconclusive}
\label{fig:3a}
\end{subfigure}
\hfill
\begin{subfigure}[b]{0.32\textwidth}
\includegraphics[width=\textwidth]{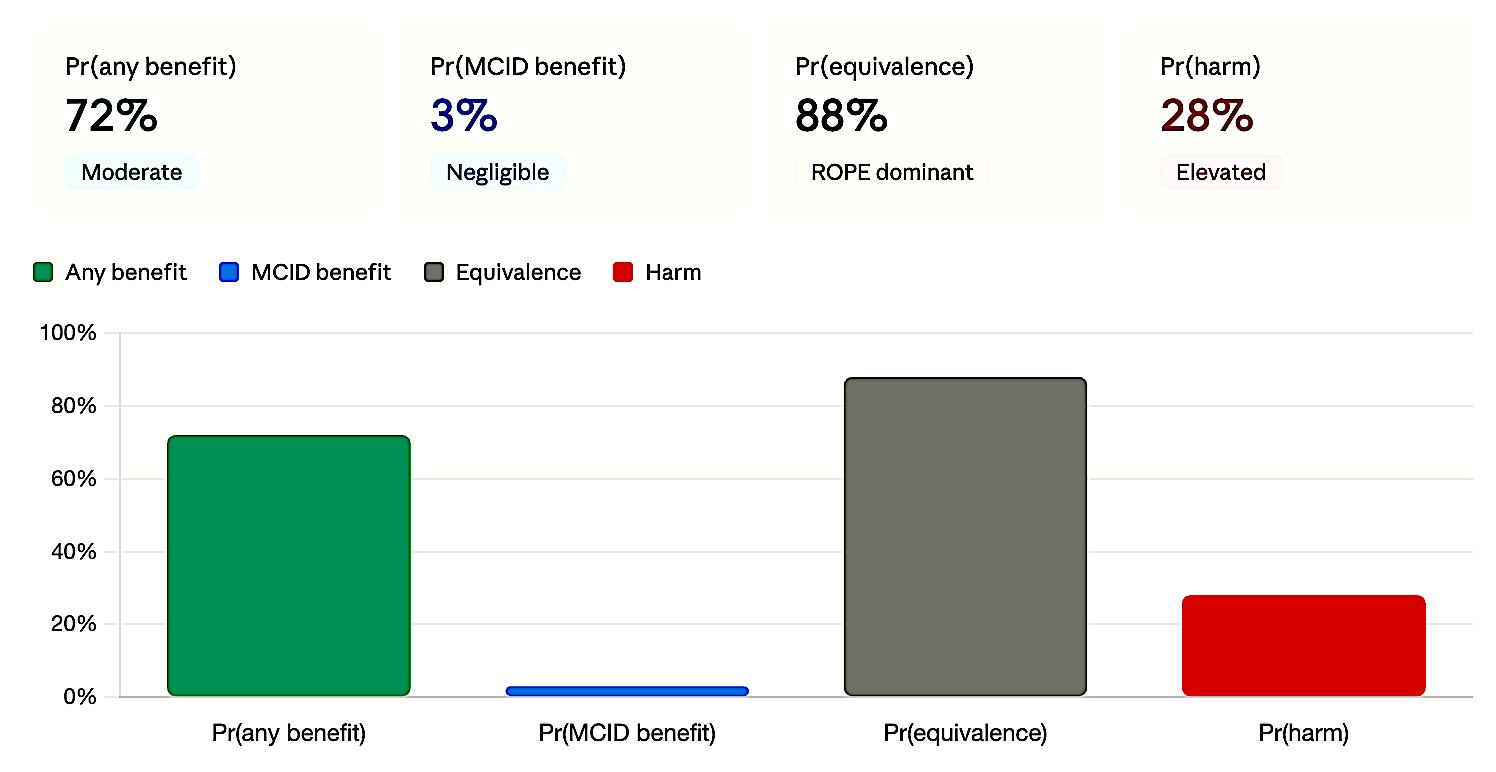}
\caption{Negative}
\label{fig:3b}
\end{subfigure}
\hfill
\begin{subfigure}[b]{0.32\textwidth}
\includegraphics[width=\textwidth]{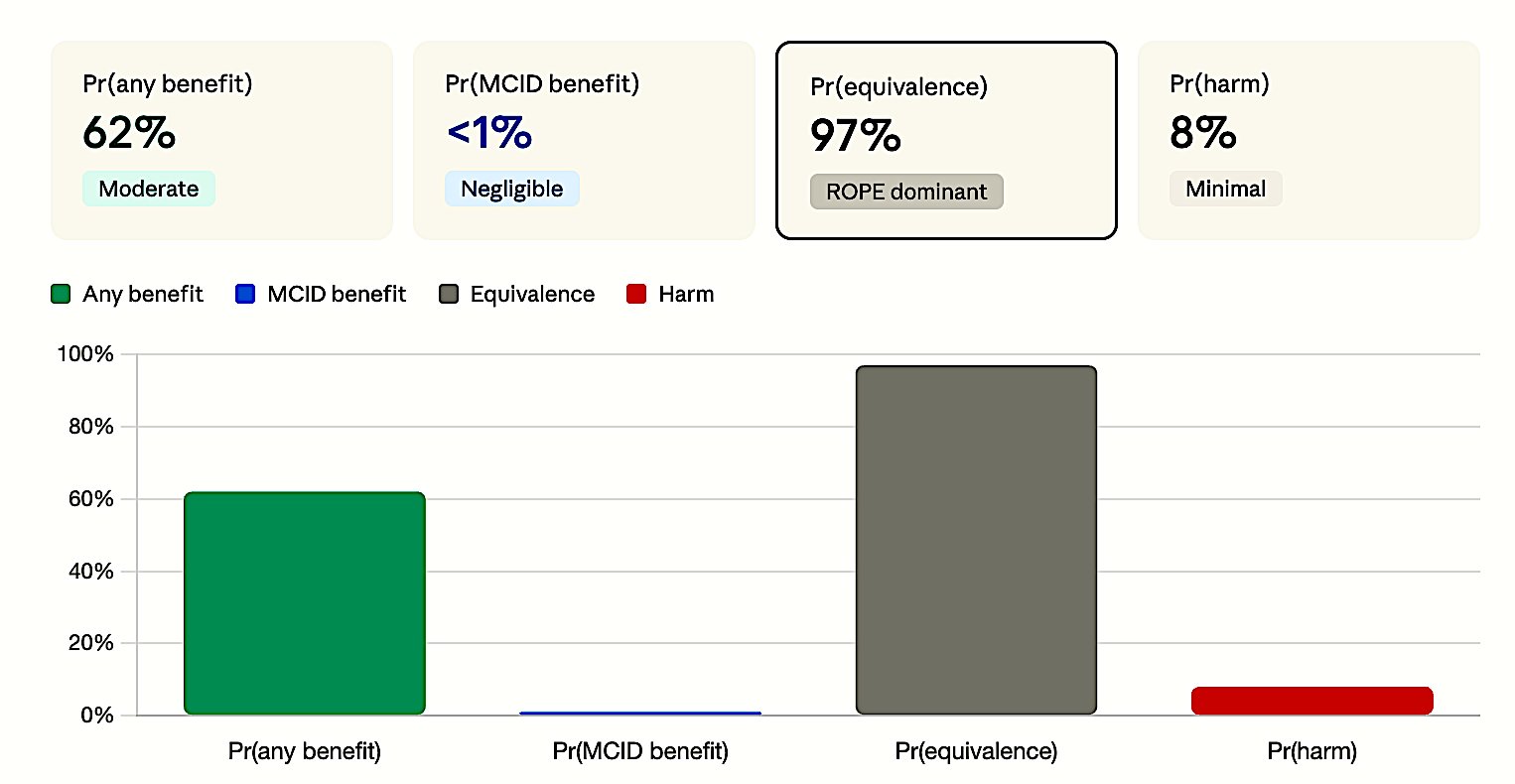}
\caption{Neutral}
\label{fig:3c}
\end{subfigure}
\caption{\textbf{The three faces of $p > 0.05$.} Three trials with non-significant $p$-values produce completely different Bayesian posterior probability profiles. (a)~Inconclusive: no metric dominates. (b)~Negative: ROPE dominates but MCID benefit $\approx 0$. (c)~Neutral: ROPE overwhelmingly dominant ($>$97\%).}
\label{fig:threefaces}
\end{figure}

%% ============================================================
\section{Bayesian Analysis for All 6 Classifications}
\label{sec:bayesian}
%% ============================================================

\subsection{Why Bayesian?}

The frequentist framework answers: ``Can we reject the null?'' (Yes/No --- same answer for all three $p > 0.05$ scenarios above). The Bayesian framework answers four separate questions: ``What is the probability of benefit? Of meaningful benefit? Of equivalence? Of harm?'' --- producing four numbers that create completely different profiles for each class.

\subsection{Three Bayesian Metrics from Zampieri et al.\ (AJRCCM 2021)}

Zampieri, Casey, Shankar-Hari, Harrell, and Harhay formalized a standardized framework for Bayesian reanalysis of clinical trials \citep{zampieri2021}:

% ==== TABLE 2: Bayesian Metrics ====
\begin{table}[!htbp]
\centering
\caption{\textbf{Bayesian posterior metrics} for trial classification (per Zampieri et al.\ 2021).}
\label{tab:metrics}
\begin{tabular}{@{}lll@{}}
\toprule
\textbf{Metric} & \textbf{Definition} & \textbf{Maps to} \\
\midrule
$\Pr(\text{outstanding benefit})$ & $\Pr(\text{OR} < \text{MCID})$ & Positive end \\
$\Pr(\text{ROPE})$ & $\Pr(1/1.1 < \text{OR} < 1.1)$ & Neutral (equivalence) \\
$\Pr(\text{severe harm})$ & $\Pr(\text{OR} > \text{MCID-harm})$ & Harmful end \\
\midrule
\multicolumn{3}{@{}l}{\textit{Supplementary metrics:}} \\
$\Pr(\text{any benefit})$ & $\Pr(\text{OR} < 1.0)$ & \\
Posterior median OR & Median + 95\% CrI & \\
Absolute risk reduction & ARR $= \text{CER} \times (1 - \text{OR})$ & \\
\bottomrule
\end{tabular}
\end{table}

These three metrics together produce a unique signature for each of the 6 classifications. The ROPE concept \citep{kruschke2018} provides the Bayesian analog to MCID --- if the Highest Density Interval falls entirely within ROPE $\rightarrow$ equivalence; outside $\rightarrow$ effect present; partial overlap $\rightarrow$ inconclusive.

\subsection{Complete Bayesian Fingerprints}

% ==== TABLE 5: Complete Bayesian Fingerprints ====
\begin{table}[!htbp]
\centering
\caption{\textbf{Complete Bayesian fingerprints} for all six trial classifications. Each class produces a distinctive posterior probability profile.}
\label{tab:fingerprints}
\small
\begin{tabular}{@{}lccccp{4.5cm}@{}}
\toprule
\textbf{Class} & \textbf{Pr(any benefit)} & \textbf{Pr(MCID benefit)} & \textbf{Pr(ROPE)} & \textbf{Pr(harm)} & \textbf{Dominant signal} \\
\midrule
Positive     & $>$99\% & $>$90\% & $<$1\% & $<$1\% & MCID benefit overwhelms \\
Imprecise~(+) & $\sim$97\% & 50--70\% & $\sim$8\% & $\sim$3\% & Gap: any benefit $\uparrow$ vs MCID benefit $\downarrow$ \\
Neutral      & $\sim$62\% & $<$1\% & $>$90\% & $\sim$5\% & ROPE (equivalence) dominates \\
Inconclusive & $\sim$74\% & $\sim$38\% & $\sim$35\% & $\sim$18\% & Nothing dominates \\
Negative     & $\sim$72\% & $\sim$3\% & $\sim$88\% & $\sim$20\% & MCID benefit $\approx 0$ + high ROPE \\
Harmful      & $<$1\% & $<$1\% & $\sim$4\% & $>$95\% & Harm dominates \\
\bottomrule
\end{tabular}
\end{table}

% ==== TABLE 3: Bayesian Classification Criteria ====
\begin{table}[!htbp]
\centering
\caption{\textbf{Bayesian classification criteria.} Primary and supporting conditions for assigning each verdict.}
\label{tab:bayesian_criteria}
\small
\begin{tabular}{@{}lp{5cm}p{5.5cm}@{}}
\toprule
\textbf{Verdict} & \textbf{Primary signal} & \textbf{Supporting condition} \\
\midrule
Positive & $\Pr(\text{outstanding benefit}) > 80\%$; $\Pr(\text{ROPE}) = 0$; $\Pr(\text{harm}) = 0$ & Harm posterior near zero across all prior specs \\
Imprecise~(+) & $\Pr(\text{any benefit})$ high; $\Pr(\text{outstanding benefit})$ 40--70\% & Gap between ``any benefit'' and ``outstanding benefit'' = diagnostic imprecision \\
Neutral & $\Pr(\text{ROPE}) > 90\%$; $\Pr(\text{outstanding benefit}) = 0$ & $\Pr(\text{severe harm}) < 10\%$; benefit effectively excluded \\
Inconclusive & All posteriors $< 50\%$; no single hypothesis wins & Posterior mass spread across benefit, ROPE, harm simultaneously \\
Negative & $\Pr(\text{outstanding benefit}) = 0$; $\Pr(\text{ROPE}) > 80\%$ & $\Pr(\text{severe harm}) < 20\%$; harm not dominant driver \\
Harmful & $\Pr(\text{severe harm}) > 40\%$; robust across prior specs & $\Pr(\text{benefit}) = 0$; harm signal prior-independent \\
\bottomrule
\end{tabular}
\end{table}

%% ============================================================
\section{Real-World Examples: Bayesian Reanalysis in Action}
\label{sec:examples}
%% ============================================================

\subsection{EOLIA --- ECMO for Severe ARDS (NEJM 2018) --- Bayesian Rescues Benefit}

\textbf{Frequentist verdict:} ``NEGATIVE'' --- 60-day mortality: ECMO 35\% vs control 46\%, RR 0.76 (95\% CI 0.55--1.04), $p = 0.09$.

Conclusion in the paper: ``Early ECMO was not associated with mortality that was significantly lower'' \citep{combes2018}.

\textbf{Bayesian reanalysis} (Goligher et al., \textit{JAMA} 2018) \citep{goligher2018}: Even under strong skepticism (equivalent to a hypothetical 264-patient trial finding zero effect), there is an 88\% probability ECMO reduces mortality. With prior studies incorporated, $\Pr(\text{benefit})$ reaches 99\%.

The $p = 0.09$ label of ``negative'' obscured what was actually strong evidence of benefit --- an 11\% absolute mortality reduction with 96\% posterior probability.

\begin{figure}[!htbp]
\centering
\includegraphics[width=0.95\textwidth]{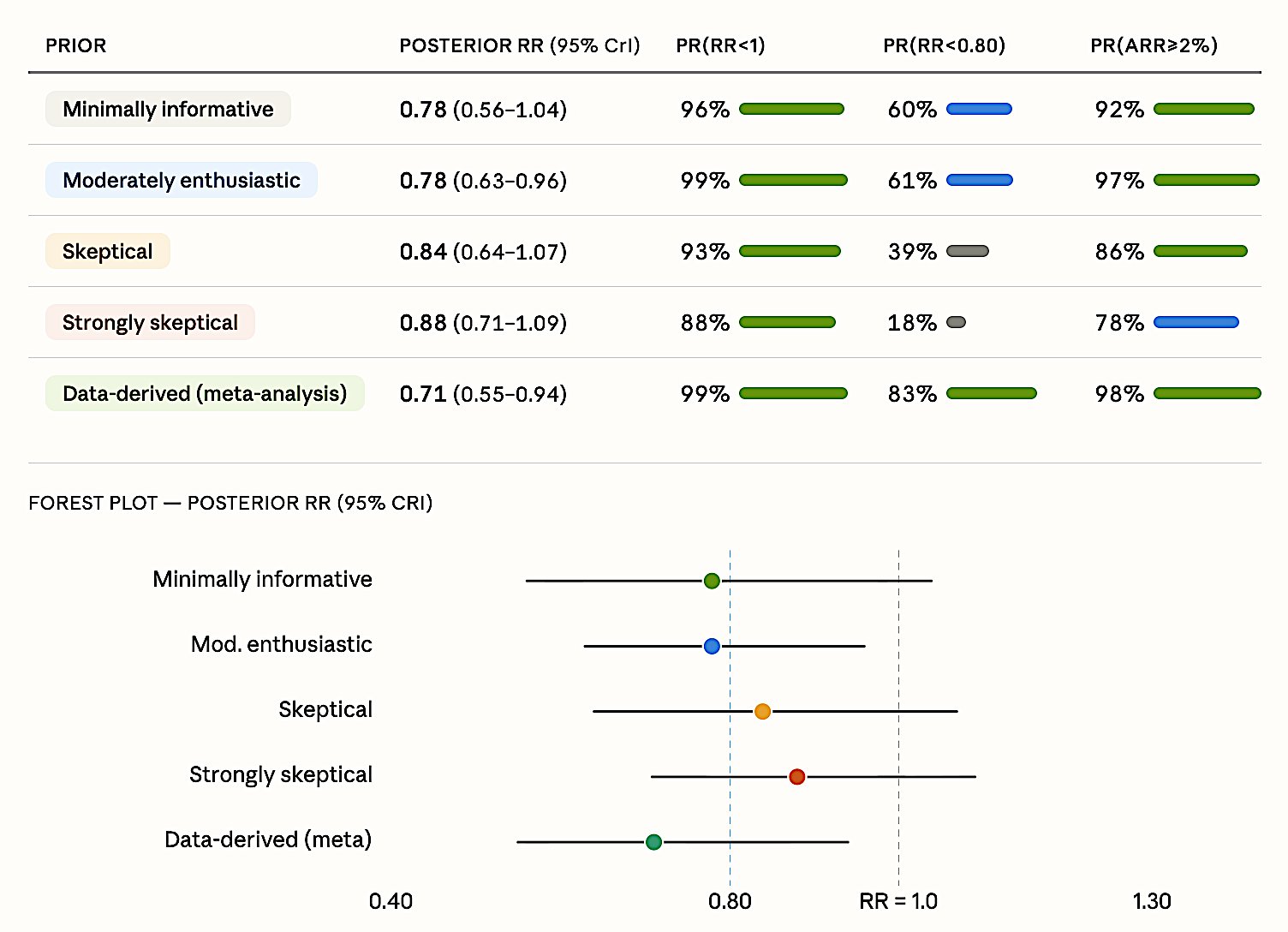}
\caption{\textbf{EOLIA Bayesian reanalysis.} Posterior RR estimates across five prior specifications. Even the strongly skeptical prior yields 88\% probability of benefit. $\Pr(\text{RR}<0.80) = 18$--$83\%$ depending on prior; $\Pr(\text{ARR}>2\%) = 78$--$98\%$.}
\label{fig:eolia}
\end{figure}

\subsection{ANDROMEDA-SHOCK --- CRT vs Lactate-Guided Resuscitation (JAMA 2019) --- Bayesian Rescues Benefit}

\textbf{Frequentist verdict:} ``NEGATIVE'' --- 28-day mortality: CRT-guided 34.9\% vs lactate-guided 43.4\%, HR 0.75 (95\% CI 0.55--1.02), $p = 0.06$ \citep{hernandez2019}.

\textbf{Bayesian reanalysis} (Zampieri et al., \textit{AJRCCM} 2020) \citep{zampieri2020}: $\Pr(\text{benefit})$ exceeds 90\% under ALL four priors --- including the pessimistic one that assumes the treatment is harmful.

A critical finding: simply switching from Cox to logistic regression yields $p = 0.022$ --- the ``negative'' label was an artifact of the statistical model choice, not the data. The Bayesian approach is immune to this model dependency.

\begin{figure}[!htbp]
\centering
\includegraphics[width=0.95\textwidth]{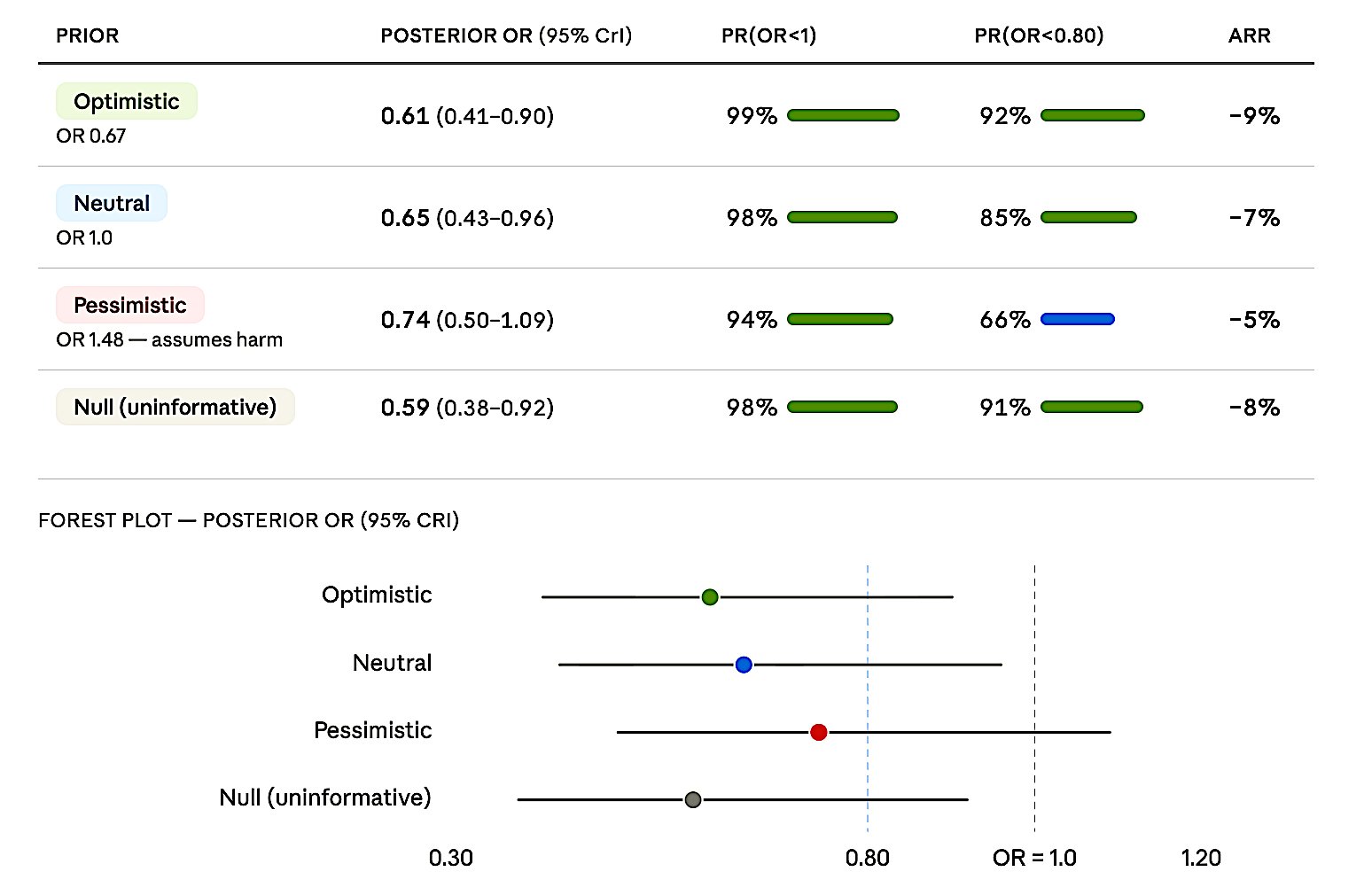}
\caption{\textbf{ANDROMEDA-SHOCK Bayesian reanalysis.} Posterior OR estimates under four priors. $\Pr(\text{OR}<1)$ ranges from 94\% (pessimistic) to 99\% (optimistic). Even the pessimistic prior yields 66\% probability of $\Pr(\text{OR}<0.80)$.}
\label{fig:andromeda}
\end{figure}

\subsection{ART --- Open-Lung Ventilation in ARDS (JAMA 2017) --- Bayesian Confirms Harm}
\label{sec:art}

\textbf{Frequentist:} borderline --- OR 1.27 (95\% CI 0.99--1.63), $p = 0.057$ \citep{cavalcanti2017}.

\textbf{Bayesian reanalysis} (Zampieri et al., \textit{AJRCCM} 2021) \citep{zampieri2021}: Even the optimistic prior (which assumes the treatment is beneficial!) yields 93.6\% probability of harm and 34.8\% probability of severe harm. $\Pr(\text{benefit}) \approx 0\%$ under ALL priors. Prior sensitivity $I^2 = 0.11$ --- priors barely matter, the data overwhelm them.

This is what a decisive HARMFUL classification looks like in Bayesian language --- the mirror image of EOLIA and ANDROMEDA-SHOCK.

\begin{figure}[!htbp]
\centering
\includegraphics[width=0.85\textwidth]{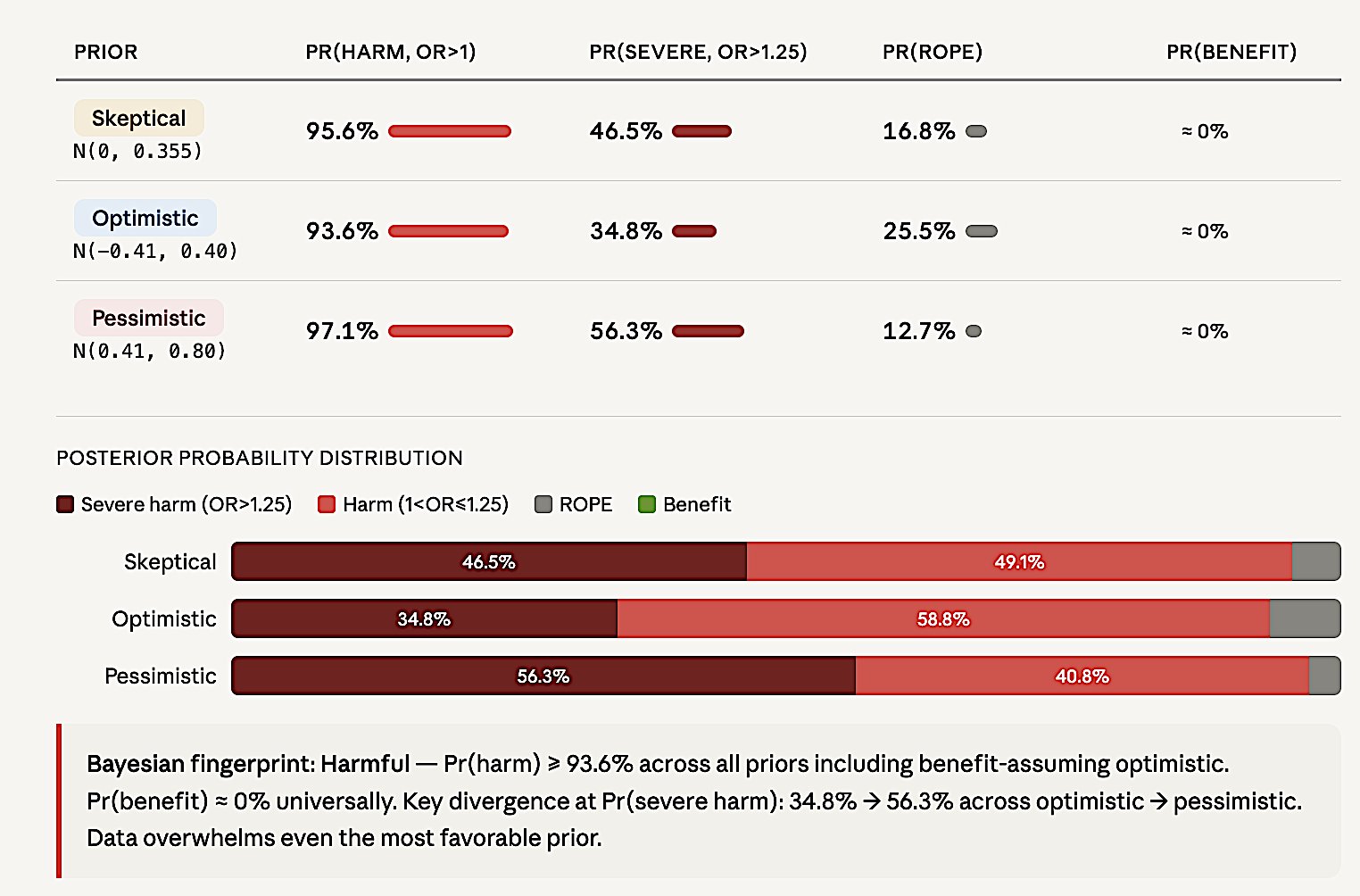}
\caption{\textbf{ART Bayesian reanalysis.} Posterior probability distributions under three priors. $\Pr(\text{harm}) > 93.6\%$ across all priors including benefit-assuming optimistic. $\Pr(\text{benefit}) = 0\%$ universally.}
\label{fig:art}
\end{figure}

\subsection{The Pattern Across All Three Reanalyses}

\textbf{Key principle:} When the conclusion is robust across skeptical-to-enthusiastic priors, it's the data talking, not the prior. Bayesian analysis works in all directions --- rescues benefit when it exists, confirms harm when it exists, and formally quantifies inconclusive when data are insufficient.

\begin{figure}[!htbp]
\centering
\includegraphics[width=0.95\textwidth]{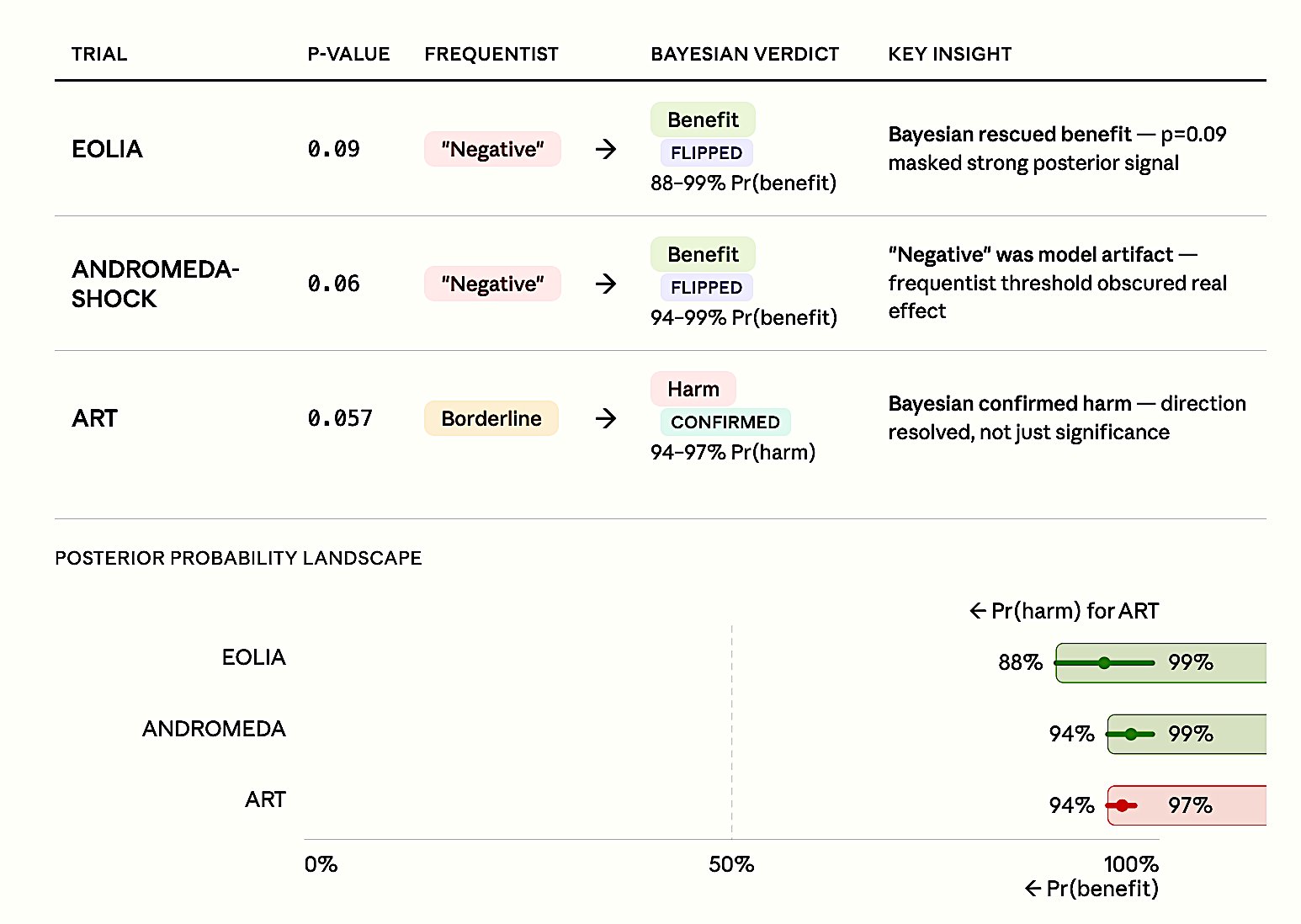}
\caption{\textbf{Pattern across three landmark reanalyses.} EOLIA and ANDROMEDA-SHOCK: frequentist ``negative'' labels flipped to benefit (88--99\% posterior probability). ART: borderline frequentist result confirmed as harmful (94--97\% posterior probability). In all three cases, $p$-values near 0.05 obscured decisive evidence that Bayesian analysis revealed.}
\label{fig:pattern}
\end{figure}

%% ============================================================
\section{Cardiology RCT Examples}
%% ============================================================

Below are analyses of selected cardiology RCTs demonstrating each of the six classifications.

% ---- Figure 8: Cardiology examples ----
\begin{figure}[!htbp]
\centering
\begin{subfigure}[b]{0.48\textwidth}
\includegraphics[width=\textwidth]{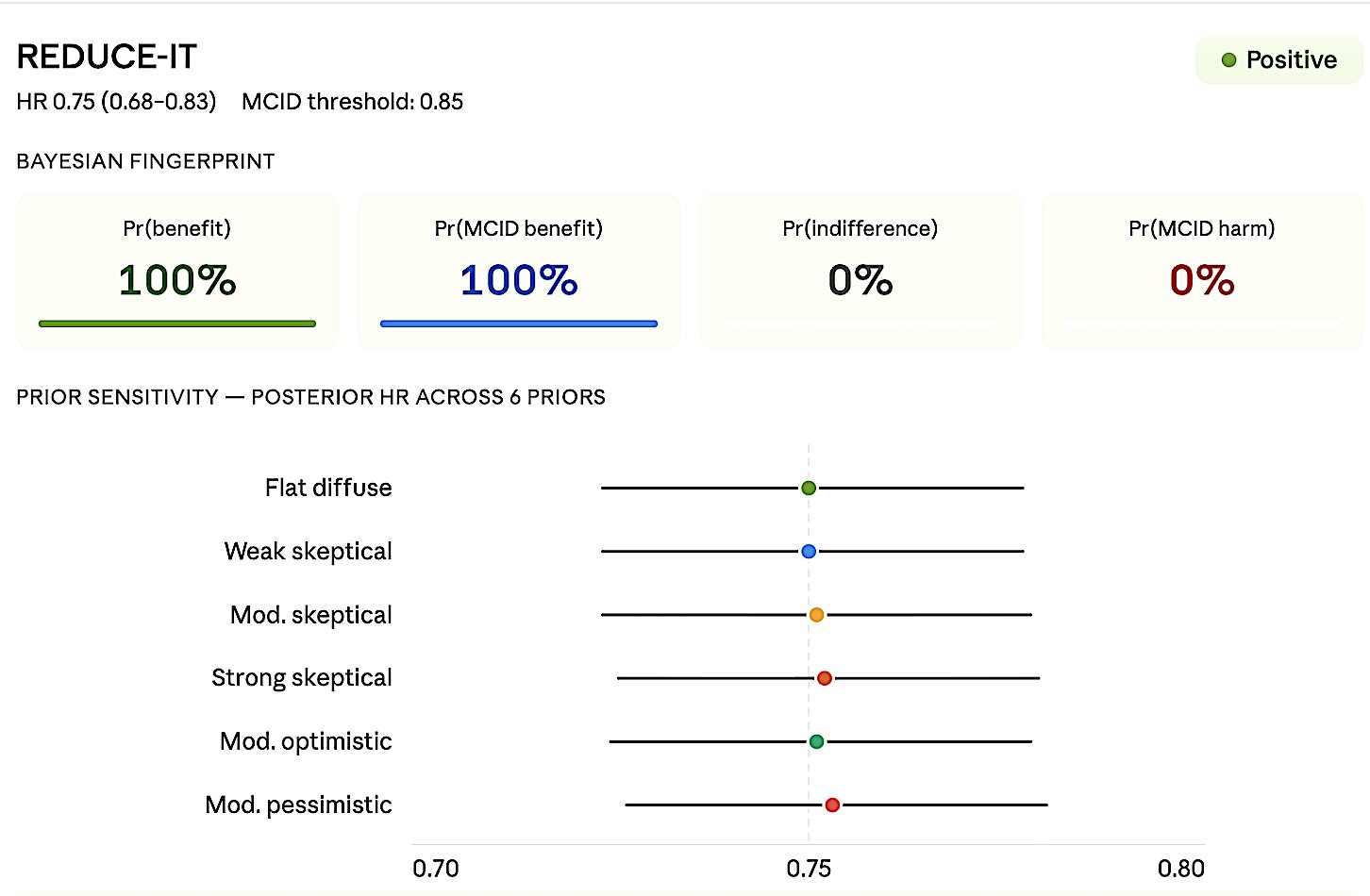}
\caption{REDUCE-IT --- \textsc{Positive}: HR 0.75 (0.68--0.83), MCID 0.85}
\label{fig:reduceit}
\end{subfigure}
\hfill
\begin{subfigure}[b]{0.48\textwidth}
\includegraphics[width=\textwidth]{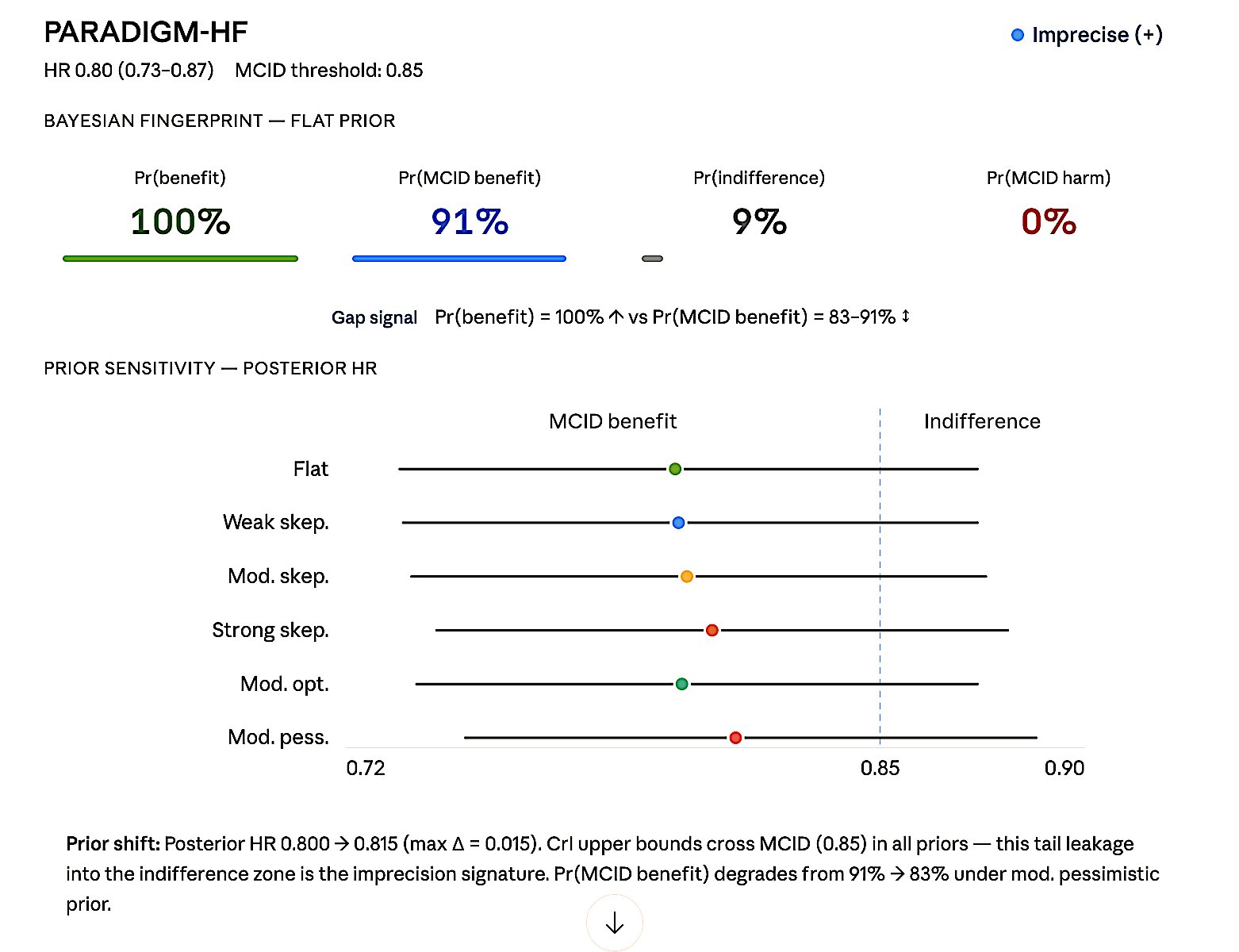}
\caption{PARADIGM-HF --- \textsc{Imprecise~(+)}: HR 0.80 (0.73--0.87), MCID 0.85}
\label{fig:paradigm}
\end{subfigure}

\vspace{0.5cm}
\begin{subfigure}[b]{0.48\textwidth}
\includegraphics[width=\textwidth]{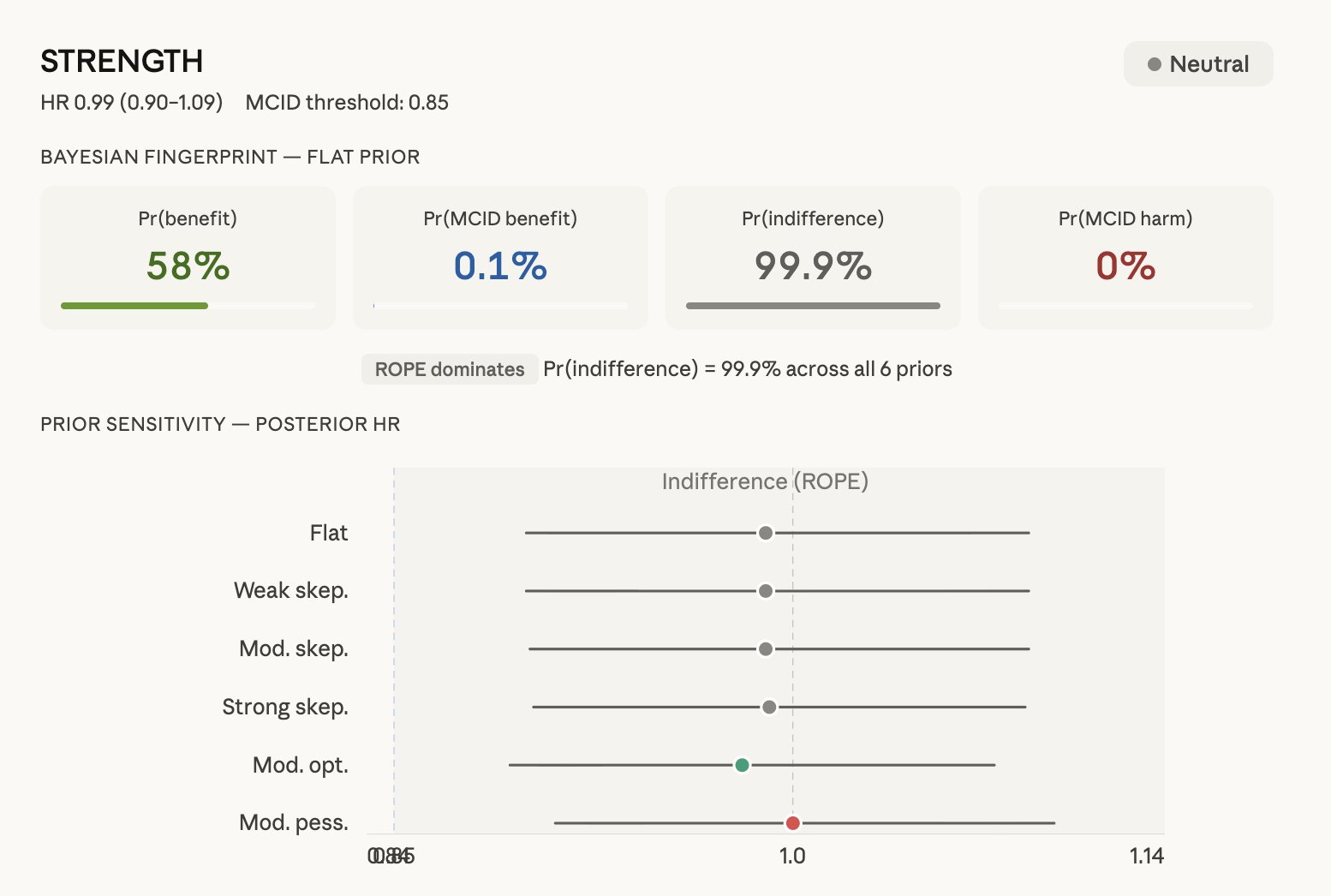}
\caption{STRENGTH --- \textsc{Neutral}: HR 0.99 (0.90--1.09), MCID 0.85}
\label{fig:strength}
\end{subfigure}
\hfill
\begin{subfigure}[b]{0.48\textwidth}
\includegraphics[width=\textwidth]{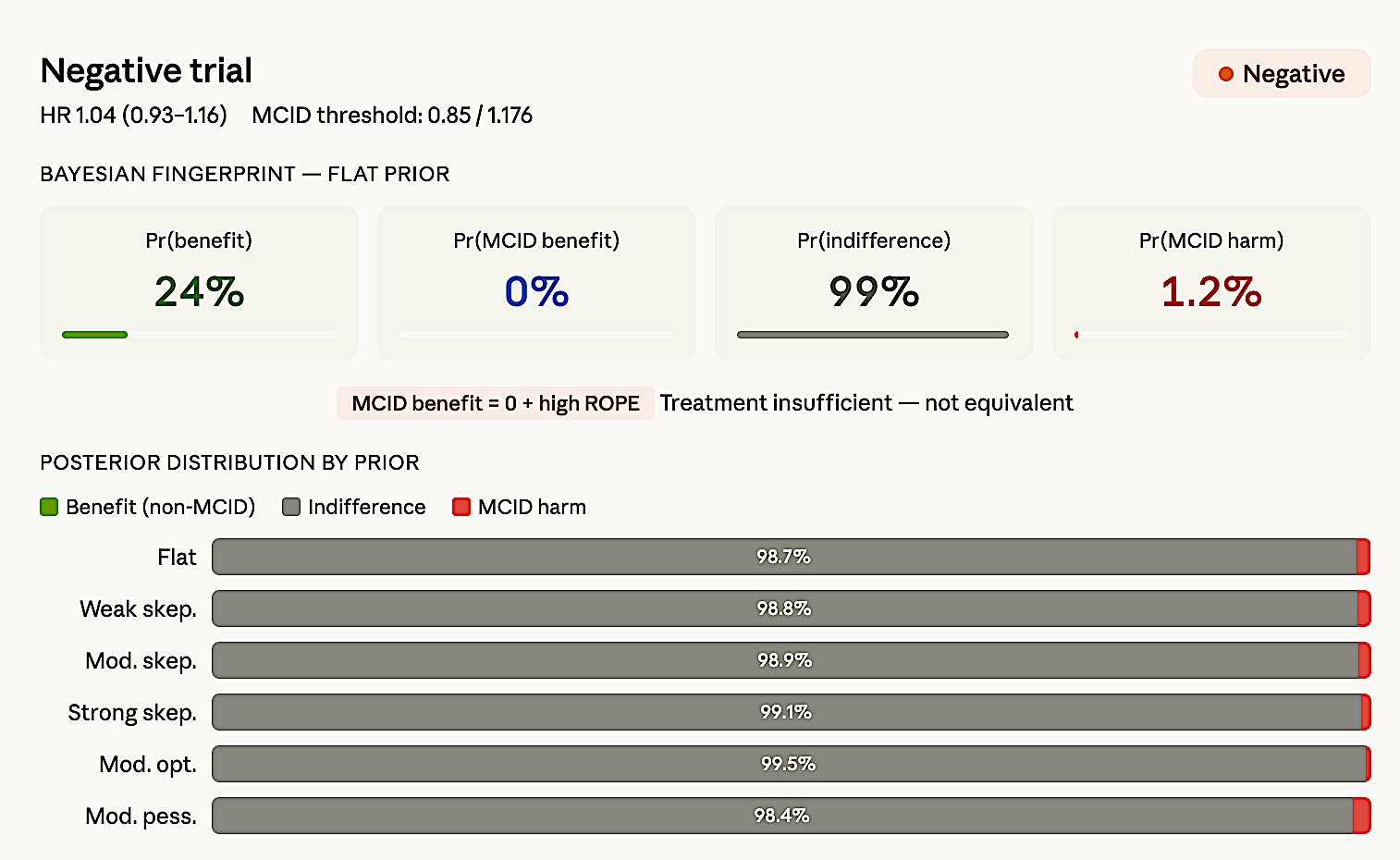}
\caption{dal-OUTCOMES --- \textsc{Negative}: HR 1.04 (0.93--1.16), MCID 0.85}
\label{fig:daloutcomes}
\end{subfigure}

\vspace{0.5cm}
\begin{subfigure}[b]{0.48\textwidth}
\includegraphics[width=\textwidth]{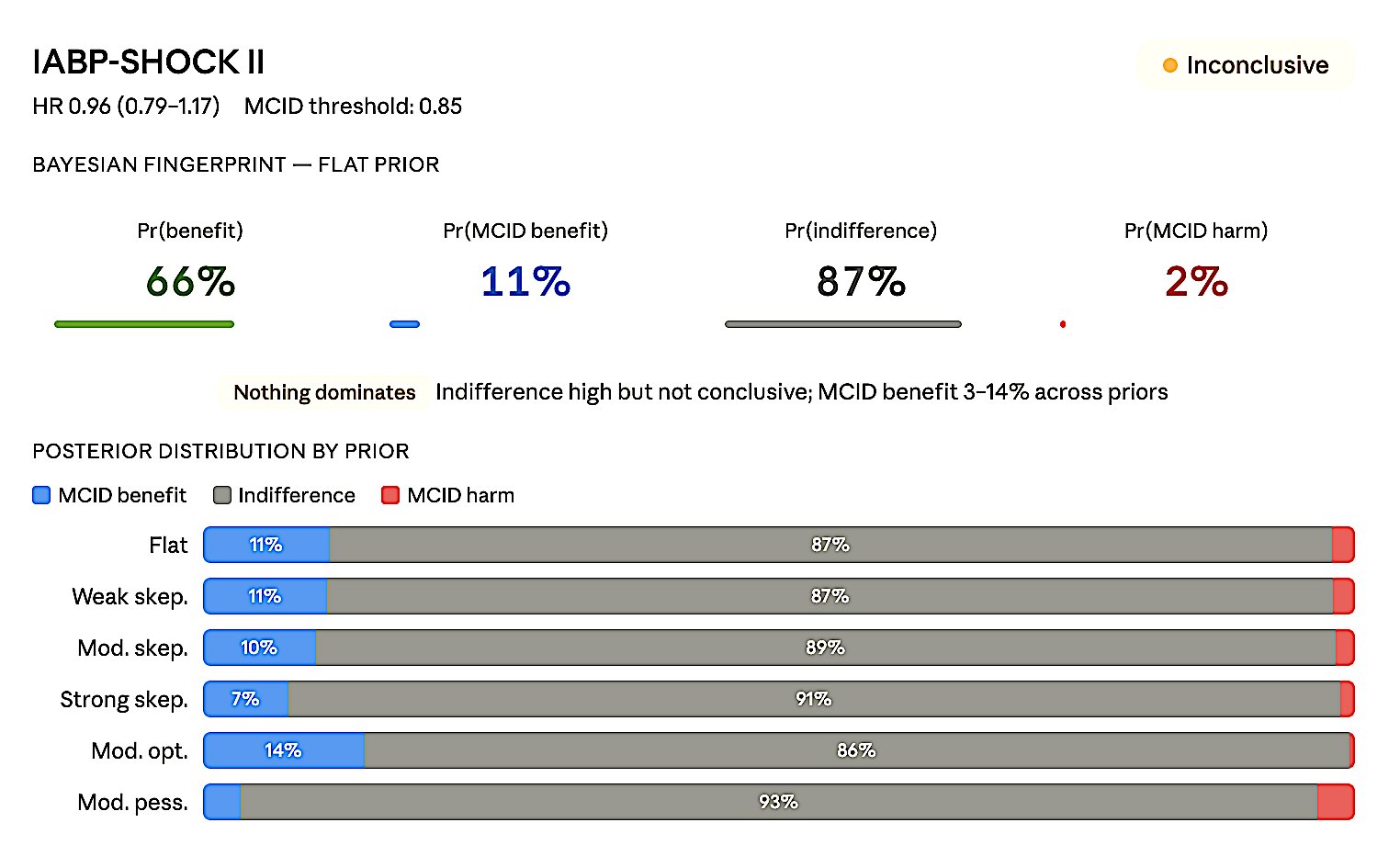}
\caption{IABP-SHOCK~II --- \textsc{Inconclusive}: HR 0.96 (0.79--1.17), MCID 0.85}
\label{fig:iabpshock}
\end{subfigure}
\hfill
\begin{subfigure}[b]{0.48\textwidth}
\includegraphics[width=\textwidth]{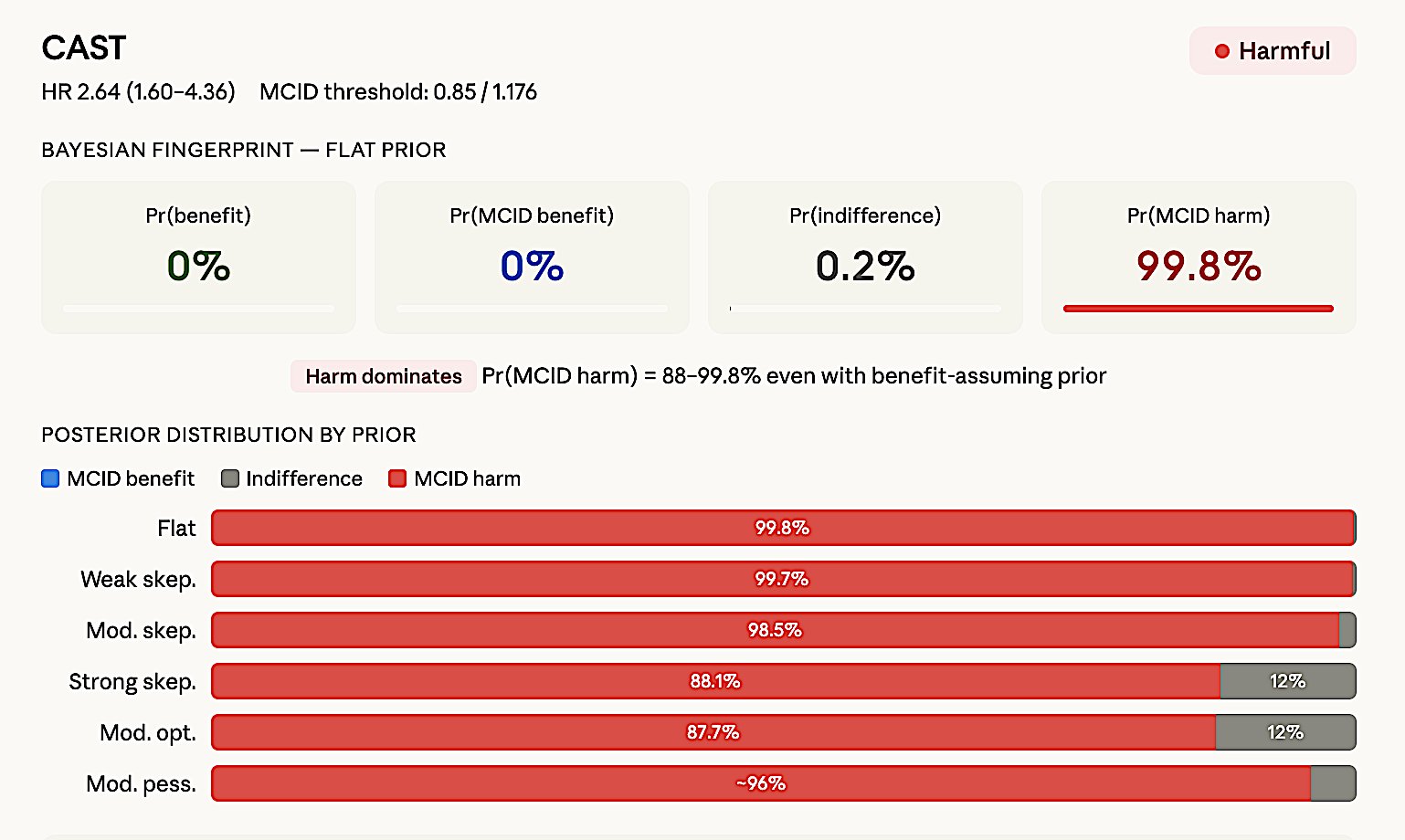}
\caption{CAST --- \textsc{Harmful}: HR 2.64 (1.60--4.36)}
\label{fig:cast}
\end{subfigure}
\caption{\textbf{Cardiology RCT examples} illustrating all six classifications with Bayesian fingerprints and prior sensitivity analyses.}
\label{fig:cardiology}
\end{figure}

%% ============================================================
\section{Most Common Errors}
%% ============================================================

\textbf{Error \#1: $p > 0.05$ = ``No difference.''} $p > 0.05$ only means the null cannot be rejected. Among 423 ``negative'' oncology RCTs: only 10.6\% had adequate power.

\textbf{Error \#2: Post-hoc power.} Post-hoc power $= f(p\text{-value}) =$ zero information. CI tells you everything.

\textbf{Error \#3: ``Neutral'' for inconclusive.} Neutral = narrow CI + MCID excluded. Wide CI = inconclusive.

\textbf{Error \#4: Taking underpowered ``positive'' at face value.} Winner's curse: effect inflated 5--$10\times$.

\textbf{Error \#5: Labeling ``negative'' when Bayesian shows strong benefit.} EOLIA ($p=0.09$, $\Pr(\text{benefit})=96\%$) and ANDROMEDA-SHOCK ($p=0.06$, $\Pr(\text{benefit})=98\%$) --- frequentist ``non-significance'' masked decisive evidence.

%% ============================================================
\section{Reporting Templates}
%% ============================================================

% ==== TABLE 6: Reporting Templates ====
\begin{table}[!htbp]
\centering
\caption{\textbf{Recommended reporting templates} for each trial classification.}
\label{tab:reporting}
\small
\begin{tabular}{@{}lcp{9cm}@{}}
\toprule
\textbf{Verdict} & \textbf{Example} & \textbf{Reporting template} \\
\midrule
Positive & \makecell{HR $= 0.74$\\(0.65--0.85)} & $>$20\% reduction threshold exceeded. Bayesian $\Pr(\text{MCID benefit}) = 97\%$. MCID benefit dominant; CI excludes null; high precision. \\
\midrule
Imprecise~(+) & \makecell{HR $= 0.70$\\(0.49--0.98)} & Benefit likely but magnitude uncertain. Bayesian $\Pr(\text{MCID benefit}) = 62\%$. Any-benefit high, MCID uncertain; wide CI; needs larger $N$. \\
\midrule
Neutral & \makecell{RR $= 0.98$\\(0.94--1.02)} & Clinically meaningful difference excluded. Bayesian $\Pr(\text{ROPE}) = 97\%$. Equivalence established; tight CI around null; high precision. \\
\midrule
Inconclusive & \makecell{HR $= 0.90$\\(0.65--1.24)} & Neither benefit nor harm excluded. Bayesian: no $\Pr$ exceeds 50\%. Nothing dominates; CI spans null widely; uninformative. \\
\midrule
Negative & \makecell{RR $= 0.95$\\(0.85--1.07)} & $>$20\% reduction excluded. Bayesian $\Pr(\text{MCID benefit}) = 3\%$. MCID benefit $\approx 0$; high ROPE; treatment insufficient. \\
\midrule
Harmful & \makecell{OR $= 1.27$\\(0.99--1.63)} & Bayesian $\Pr(\text{severe harm}) = 47\%$ even with optimistic prior. Harm dominates; prior-resistant signal; safety concern. \\
\bottomrule
\end{tabular}
\end{table}

%% ============================================================
\section{Non-Inferiority and Equivalence}
%% ============================================================

\textbf{Non-inferiority:} One-sided threshold $\Delta$. CI harm-side within $\Delta$ $\rightarrow$ ``non-inferior.'' Exceeds $\rightarrow$ ``inferior.'' Crosses $\rightarrow$ inconclusive. FDA NI guidance (2016): M1/M2 margin framework.

\textbf{Equivalence:} Two-sided $\pm\Delta$. CI within band $\rightarrow$ ``equivalent.'' Crosses $\rightarrow$ inconclusive.

%% ============================================================
\section*{Conclusion}
%% ============================================================

``Never interpret $p > 0.05$ as `no effect.' Always report CI + effect size + clinical significance. When $p$ is near 0.05, compute Bayesian posteriors before labeling the trial.'' --- Harrell, Pocock, Zampieri, ASA, ICH~E9.

%% ============================================================
%% REFERENCES
%% ============================================================
\bibliographystyle{unsrtnat}

\end{document}